\documentclass[10pt, conference, letterpaper]{IEEEtran}
\usepackage{cite}
\usepackage{amsmath}
\usepackage{amsfonts}
\usepackage{amsthm}
\usepackage[pdftex]{graphicx}
\ifCLASSOPTIONcompsoc
    \usepackage[caption=false, font=normalsize, labelfont=sf, textfont=sf]{subfig}
\else
\usepackage[caption=false, font=footnotesize]{subfig}
\fi

\theoremstyle{plain}
\newtheorem{theorem}{Theorem}
\theoremstyle{plain}
\newtheorem{corollary}{Corollary}
\theoremstyle{plain}
\newtheorem{lemma}{Lemma}
\theoremstyle{definition}
\newtheorem{definition}{Definition}
\theoremstyle{remark}
\newtheorem*{remark}{Remark}
\newcommand{\isep}{\mathrel{{.}\,{.}}\nobreak}

\IEEEoverridecommandlockouts
\begin{document}
\title{Approximations and Bounds for $(n, k)$ Fork-Join Queues: A Linear Transformation Approach}

\author{
\IEEEauthorblockN{Huajin Wang\IEEEauthorrefmark{1}\IEEEauthorrefmark{2},
Jianhui Li\IEEEauthorrefmark{1},
Zhihong Shen\IEEEauthorrefmark{1} and
Yuanchun Zhou\IEEEauthorrefmark{1}
}
\IEEEauthorblockA{\IEEEauthorrefmark{1}Computer Network Information Center, Chinese Academy of Sciences, China
\\ Email: \{wanghj, lijh, bluejoe, zyc\}@cnic.cn}
\IEEEauthorblockA{\IEEEauthorrefmark{2}University of Chinese Academy of Sciences, China
}
\thanks{This work was supported by the National Key Research Program of China (2016YFB1000600, 2016YFB0501900). Corresponding author is Jianhui Li.}
}

\maketitle
\begin{abstract}
Compared to basic fork-join queues, a job in $(n,k)$ fork-join queues only needs its $k$ out of all $n$ sub-tasks to be finished. Since $(n,k)$ fork-join queues are prevalent in popular distributed systems, erasure coding based cloud storages, and modern network protocols like multipath routing, estimating the sojourn time of such queues is thus critical for the performance measurement and resource plan of computer clusters. However, the estimating keeps to be a well-known open challenge for years, and only rough bounds for a limited range of load factors have been given. In this paper, we developed a closed-form \texttt{linear transformation technique} for jointly-identical random variables: An order statistic can be represented by a linear combination of maxima. This brand-new technique is then used to transform the sojourn time of non-purging $(n,k)$ fork-join queues into a linear combination of the sojourn times of basic $(k,k), (k+1,k+1),...,(n,n)$ fork-join queues. Consequently, existing approximations for basic fork-join queues can be bridged to the approximations for non-purging $(n,k)$ fork-join queues. The uncovered approximations are then used to improve the upper bounds for purging $(n,k)$ fork-join queues. Simulation experiments show that this linear transformation approach is practiced well for moderate $n$ and relatively large $k$.
\end{abstract}


\section{Introduction}
\begin{figure*}
\centering
\subfloat[Basic $(3,3)$ Fork-Join Queues]
	{\includegraphics[width=0.7\columnwidth]{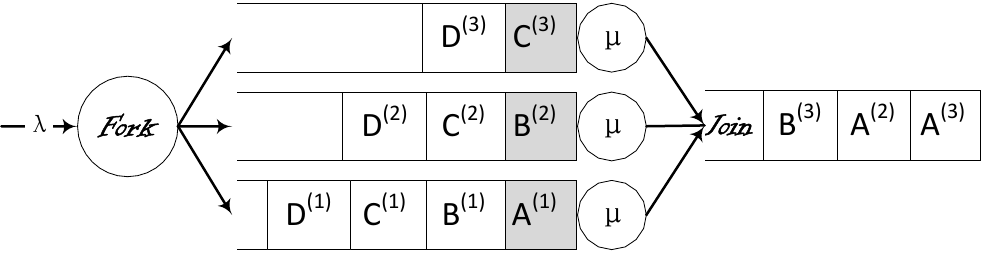}}\hfill
\subfloat[Non-Purging $(3,2)$ Fork-Join Queues]{\includegraphics[width=0.64\columnwidth]{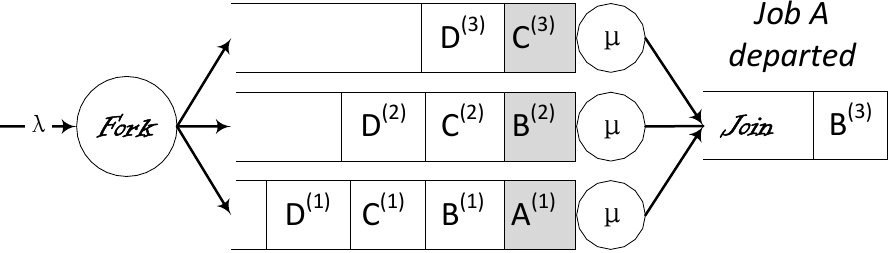}}\hfill
\subfloat[Purging $(3,2)$ Fork-Join Queues]{\includegraphics[width=0.64\columnwidth]{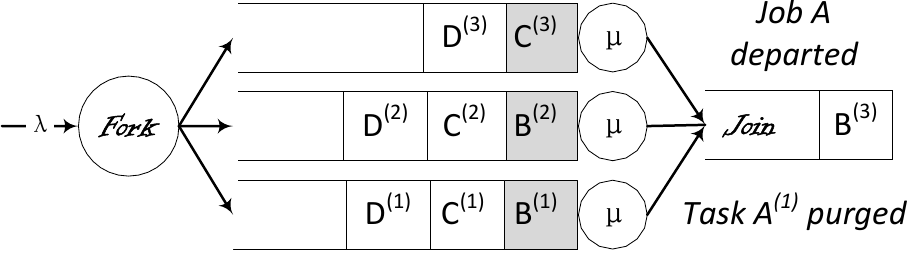}}
\caption{Fork-Join Queues: $A^{(1)}$ Is a Sub-Task of Job $A$}
\label{queues}
\end{figure*}

The performance of fork-join queues is a highly focused research topic for many years for the ubiquitousness of fork-join queues in both real-life workflows and computing systems. In a fork-join queueing system, a job is \textit{forked} into $n$ sub-tasks when it arrives at a control node, and each sub-task is sent to a single node to be conquered. Results of finished sub-tasks are summarized at a central \textit{join} node. When the job arrival rate $\lambda$ is high, a sub-task may have to wait for service in the sub-queue of its hosting node in a first-come-first-serving order. A basic fork-join queue considers a job is done after all results of the job have been received at the join node (see Fig. \ref{queues} (a)).

In Big Data era, more and more mainstream computing infrastructures become distributively deployed, and inevitably recruit fork-join queues to facilitate the storing and processing of large-scale datasets. For example: 1) Cassandra\cite{Lakshman:2010ef} and Dynamo\cite{DeCandia:2007cn}, two popular key-value data stores, use fork-join queues to concurrently perform read and write operations on all the replicas of the target key-value pairs; 2) The client of an $(n,k)$ MDS (maximum distance separable) erasure coding based cloud storage system only needs to retrieve any $k$ out of all $n$ blocks of a file to reconstruct the file; 3) The data transmission process of multipath routing protocols can generally be simplified as a multi-stage fork-join queueing process.

Latency is commonly a critical concern in building and optimizing Big Data clusters. For example, in Amazon's cloud platform, services commonly have latency requirements which are in general measured at the $99.9^{th}$ percentile of the distribution \cite{DeCandia:2007cn}. The Dynamo storage system must be capable of meeting such stringent SLAs. In this scenario, basic fork-join queues may cause serious performance issues when the number of data replicas are large, since they require all the sub-tasks of a job to be finished before making the job's response. By contrast, $(n,k)$ fork-join queues, as named in \cite{Joshi:2012cv}, only require the job's any $k$ out of $n$ sub-tasks to be finished, and thus have performance advantages in such scenarios. For example, a write request in Casandra can either be responded when a quorum of replicas have been successfully written, or just get responded once the fast answer from all touched replicas is acknowledged when there is a need to pursue high throughputs.

As depicted in Fig. \ref{queues}, there are mainly two versions of $(n,k)$ fork-join queues: The purging one removes all the remaining sub-tasks of a job from both sub-queues and service stations once it receives the job's $k^{th}$ answer. The file retrieval process from an MDS coded cloud storage system is such an example.
As a contrast, the non-purging one keeps queuing and executing remaining sub-tasks. For example, a write operation in Cassandra needs to update all the replicas of the target key-value pair, while can response to user as soon as a quorum of replicas have been successfully written.

\paragraph{The State-of-the-Art Research on Basic Fork-Join Queues}
 The popularity of fork-join systems has drawn great attentions from database/OS/networking communities to the performance analyses of fork-join queues for a rather long period of time. Unfortunately, there is still no exact closed-form solution of the sojourn time of the job in $n\geq 3$ basic fork-join queues. The difficulty lies in the fact that the sojourn times of a job's sub-tasks are not independent, as their hosting sub-queues share the same sub-task arrival process. Since most of existing exact analysis techniques are developed for independent and identical (iid) random variables, it is very hard to trace the sojourn time distribution for fork-join queues.

For $n\geq 3$ fork-join queues under Poisson job arrival process and with iid exponential service time distributions, Nelson et al. \cite{Nelson:1988jk} proposed an initiative approximation technique which is based on the fact that the sojourn times $X_{1\isep k}$ of sub-tasks $1,2,...,k$ are associated variables, whose maximum can be bounded by the maximum of their iid equivalents \cite{Esary:1967eo}: $P(X_{1\isep n} \leq t)\geq\prod_{i=1}^{n}P(X^{IID}_i\leq t)$. According to that, the upper bounds and closed-form approximations of the sojourn time were given in this work. Simulation experiments in \cite{Lebrecht:2007tm} showed that Nelson's approximation is still the most reliable one, compared to following works such as \cite{Varma:1994gm} and \cite{Varki:2001wc}.

\paragraph{The State-of-the-Art Research and Open Challenges on $(n,k)$ Fork-Join Queues}

Despite the popularity of $(n,k)$ fork-join queues in Big Data systems and many other fields, there are even no practical approximations on the sojourn time of $(n,k)$ fork-join queues: Unlike the maximum, the $k^{th}$ order statistic cannot be bounded by using associated variables' property, which makes the sojourn time of $(n,k)$ fork-join queues more hard to analyze, compared to basic fork-join queues.

\begin{figure}[h!]
\centering
\includegraphics[width=0.8\columnwidth]{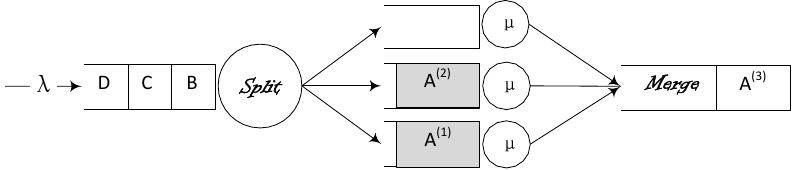}
\caption{A $(3,2)$ Split-Merge Queue}
\label{sm-queue}
\end{figure}

Currently, there are only exact quantity analyses for purging $(n,1)$ fork-join queues \cite{Gardner:2015kb,Lee:2017gi}, because such a queue is equivalent to a single queue with $n$ times the sub-queue's service rate. For general purging $(n,k)$ fork-join queues, there are only rough bounds have been given: Joshi et al. \cite{Joshi:2012cv,Joshi:2017bj} resort to the split-merge queue model
(see Fig. \ref{sm-queue}) 
 to find proper upper and lower bounds. Compared to purging $(n,k)$ fork-join queues, all empty sub-queues in the split-merge model are blocked and cannot serve subsequent tasks until $k$ sub-tasks of the current job are completed, which makes the split-merge model much easier to trace. However, these split-merge based bounds tend to be extremely loose when increasing $k$ or the load factor $\rho$, as we depict in Section \ref{bounds}.

Since non-purging $(n,k)$ fork-join queues cannot be reduced to the split-merge model, they are more difficult to analyze, even including $(n,1)$ queues. Recently, Fidler et al. \cite{Fidler:2016tw} gave non-asymptotic statistical bounds on the sojourn times for non-purging fork-join queues. However, no reasonable approximations have been proposed.

\paragraph{Methodology and Contributions}
This paper aims at fixing the lack of proper approximations for non-purging $(n,k)$ fork-join queues and tackling the uncontrollability of bounds for purging $(n,k)$ fork-join queues. To achieve these objectives, we trace fork-join queues in a fundamental way: The linear relationship between $(n,k)$ fork-join queues and their basic $(k,k), (k+1,k+1),...,(n,n)$ equivalents is depicted for the first time; This relationship is then used to bridge the existing approximations for basic fork-join queues to the approximations and bounds for $(n,k)$ fork-join queues.

Our innovations and contributions are highlighted as follows:
\begin{itemize}
  \item A brand-new closed-form \texttt{linear transformation technique} for jointly-identical random variables, by which order statistics can be transformed into a closed-form linear combination of maxima. Besides, there is no need to assume the independence of variables.
  \item The first reasonable and practical method to approximate the expected sojourn time of non-purging $(n,k)$ fork-join queues with general service time distributions. This method relies on the cooperation between the linear transformation technique and the existing approximations for basic fork-join queues.
  \item Improvements over the upper bounds on the expected sojourn time of purging $(n,k)$ fork-join queues, which are gained by resorting the bounds to that of the non-purging equivalent $(n,k)$ fork-join queues.
\end{itemize}

This paper is organized as follows: The linear transformation technique is developed in Section \ref{pre}; This technique is then employed in Section \ref{app} to find proper approximations for non-purging $(n,k)$ fork-join queues; The flaws of existing bounds for purging $(n,k)$ fork-join queues and our improvements over upper bounds are depicted in Section \ref{bounds}; In Section \ref{discuss}, we discuss the limitation of this linear transformation technique; Related works are reviewed in Section \ref{review}; We conclude this work and point out some promising future research directions in Section \ref{con}.

\section{Preliminaries: Linear Transformations of Order Statistics}\label{pre}
In this section, we consider a family of rvs (random variables) $X_1, X_2,...,X_n$ (denoted as $X_{1\isep n}$) defined on a probability space, and let $X_{(n,k)}$ denotes their $k^{th}$ order statistic, $P_k$ denotes the possibility $P(X_1\leq t, X_2\leq t,...,X_k\leq t)$ and $P_{n,k}$ denotes the possibility $P(X_1\leq t, X_2\leq t,...,X_k\leq t, X_{k+1}>t, X_{k+2}>t,...,X_n>t)$. Obviously, $P_k$ is the distribution of the maximum of $X_{1\isep k}$.

\begin{definition}[Jointly-Identical]
	For $n$ identically distributed rvs $X_{1\isep n}$ and $\forall k\in [1\isep n]$, if any $k$ arbitrarily chosen rvs  keep the same joint probability distribution, these $n$ identical rvs are named as jointly-identical rvs.
\end{definition}

\begin{lemma}{For $n$ jointly-identical rvs $X_{1\isep n}$, $$P_{n,k}=\sum_{i=k}^{n}A^{n,k}_i{P_{i}}, 1\leq \forall k \leq n,$$
 where the const coefficient $A^{n,k}_i$ can be calculated by the following recurrence: $$A^{n,k}_i=\left\{
\begin{array}{lc}
1&{i=k},\\
-\sum_{j=1}^{i-k}{n-i+j \choose j}A^{n,k}_{i-j}&{k+1\leq i\leq n}.
\end{array}
\right.$$
}\label{th-mp}
\begin{proof}
Let $P_{\overline{n-k}|k}$ denotes $P(X_{k+1}>t,X_{k+2}>t,...,X_n > t|X_1\leq t,X_2\leq t,...,X_k\leq t)$ and 
$P_{\overline{n-i},i-k|k}$ denotes $P(X_{i+1}>t,X_{i+2}>t,...,X_n > t, X_{k+1}\leq t, X_{k+2}\leq t, X_{i}\leq t|X_{1}\leq t,X_{2}\leq t,...,X_k \leq t)$, $k+1\leq i \leq n$. Certainly, we have
\begin{gather}
P_{\overline{n-k}|k}=\frac{P_{n,k}}{P_{k}}\text{ , }P_{\overline{n-i},i-k|k}=\frac{P_{n,i}}{P_{k}}\label{eq1}.
\end{gather}
As $X_{1\isep n}$ are jointly-identical rvs, the following equation holds:
\begin{gather}
P_{\overline{n-k}|k}=1-\sum_{i=k+1}^{n}{n-k \choose i-k}{P_{\overline{n-i},i-k|k}}\label{eq2}.
\end{gather}
By insertion of Eq. \ref{eq1} into Eq. \ref{eq2}, the following recurrence holds:
\begin{equation}\label{eq4}
  P_{n,k}=P_{k}-\sum_{i=k+1}^{n}{n-k \choose i-k}P_{n,i}
\end{equation}
Expanse Eq. \ref{eq4}, we complete the proof of Lemma \ref{th-mp}.
\end{proof}
\end{lemma}

\subsection{Linear Transformation of Order Statistics}

\begin{theorem}[LT of Order Statistics]{For $n$ jointly-identical rvs $X_{1\isep n}$, there exists a linear transformation from maxima to order statistics:
$$
\begin{bmatrix}
    X_{(n,1)}\\
    X_{(n,2)}\\
    \vdots \\
    X_{(n,n)}
\end{bmatrix}
=
\begin{bmatrix}
    W_1^{n,1} & W_2^{n,1} & \dots  & W_n^{n,1} \\
    0 & W_2^{n,2} & \dots  & W_n^{n,2} \\
    \vdots & \vdots & \ddots & \vdots \\
    0 & 0 & \dots  & W_n^{n,n} \\
\end{bmatrix}
\begin{bmatrix}
    X_{(1,1)}\\
    X_{(2,2)}\\
    \vdots \\
    X_{(n,n)}
\end{bmatrix}.
$$
Namely, $X_{(n,k)}=\sum_{i=k}^{n}W_i^{n,k}X_{(i,i)}, 1\leq \forall k \leq n,$ where the const coefficient
\begin{equation}
  W_i^{n,k}=\sum_{j=k}^{i}{n \choose j}A^{n,j}_i \label{eq-w-coe}
\end{equation}}\label{the2}
\begin{proof}
Let $F_{n,k}\equiv P(X_{(n,k)}\leq t)$ be the probability distribution of the $k^{th}$ order statistic. Equivalently, we need to prove
\begin{equation}
F_{n,k}=\sum_{i=k}^{n}W_i^{n,k}P(X_{(i,i)}\leq t)=\sum_{i=k}^{n}W_i^{n,k}P_{i}. 	\label{eq-fnk-proof}
\end{equation}

As $F_{n,k}=\sum_{i=k}^{n} {n \choose i}P_{n,i}$ and $P_{n,k}=\sum_{i=k}^{n}A^{n,k}_i{P_{i}}$ (Lemma \ref{th-mp}), we derive the following recurrence:
\begin{equation}\label{eqf}
F_{n,k}=\left\{
\begin{array}{lc}
P_n&{k=n},\\
F_{n,k+1}+{n \choose k}\sum_{i=k}^{n}{A^{n,k}_iP_{i}}&{1\leq k\leq n-1}.
\end{array}
\right.
\end{equation}
Expanse Eq. \ref{eqf}, we complete the proof of  Eq. \ref{eq-fnk-proof}.
\end{proof}
\begin{remark}
There is no need to assume the independence of the identical rvs $X_{1\isep n}$.
There may exist other linear transformations of order statistics than the one given by Theorem \ref{the2}.
\end{remark}
\end{theorem}
\begin{definition}[$W$ Coefficient]{For any possible linear transformations from maxima to order statistics, the corresponding const coefficient $W_i^{n,k}$ of $X_{(i,i)}$ is named as a $W$ coefficient.}
\begin{remark}
The calculation of the $W$ coefficient given by Eq. \ref{eq-w-coe} is not straightforward, as it consists of many items. We use a simple solver (see Appendix) to give $W$ coefficient values.
\end{remark}
\end{definition}

\subsection{Linear Transformation of Expectations}
 
Let $\mathbb E_{n,k}\equiv E[X_{(n,k)}]$ be the expectation of the $k^{th}$ order statistic of rvs $X_{1\isep n}$. Specially, we use $\mathbb E_{n}$ to denote $\mathbb E_{n,n}$. Then the following theorem holds.
\begin{theorem}[LT of Expectations]{For $n$ jointly-identical rvs $X_{1\isep n}$, there exists a linear transformation from the expectations of maxima to the expectations of order statistics:
\begin{equation}
  \mathbb E_{n,k}=\sum_{i=k}^{n}W_i^{n,k}\mathbb E_{i}, 1\leq \forall k \leq n.
\end{equation}}\label{the3}
\begin{proof}
$E[X_{(n,k)}]=E[\sum_{i=k}^{n}W_i^{n,k}X_{(i,i)}]=\sum_{i=k}^{n}W_i^{n,k}\mathbb E_{i}$
\end{proof}
\end{theorem}

\section{Approximations for Non-Purging $(n,k)$ Fork-Join Queues}\label{app}

We consider a homogenous cluster consisting of $n$ nodes, where each node $i$ ($1\leq i\leq n$) has the same service time distribution when processing sub-tasks of the same job. Each node owns a first-come-first-serving sub-queue $q_i$ ($1\leq i\leq n$) with the assumption of unlimited queue capacity. These $n$ sub-queues constitute a homogenous fork-join queue.

Each incoming job consists of $n$ tasks. Let $t_i^j$ be the sojourn time of the $j^{th}$ job's sub-task assigned to node $i$. Then, the stable sojourn time of a sub-task in the sub-queue $q_i$ is $t_i\equiv \lim_{j\to\infty}t_i^j$, and the sojourn time of a job in the $(n,k)$ fork-join queue is the $k^{th}$ order statistic $t_{(n,k)}$ consequently.

\begin{lemma}{For an $(n,n)$ fork-join queue, the sub-queues' stable sojourn times $t_{1\isep n}$ constitute a family of jointly-identical rvs.}\label{the4}

\begin{proof}
Recall that all the sub-queues have an unlimited capacity, then the sub-task sojourn time distribution of a sub-queue depends only on the job arrival process and the sub-queue's service time distribution.
As these sub-queues are under the same job arrival process and have the same service time distribution, the sub-queues' stable sojourn times $t_{1\isep n}$ are identical rvs.

By the constitution methodology of fork-join queues, an $(n,n)$ fork-join queue is symmetrical, which means its sub-queues are interchangeable and thus any $k$ arbitrarily chosen sub-queues keep the same joint probability distribution. By definition the stable sojourn times $t_{1\isep n}$ are jointly-identical rvs.
\end{proof}	
\end{lemma}

\begin{definition}[$(\lambda,\mu)$-Equivalent Queues]{Those basic fork-join queues, purging/non-purging $(n,k)$ fork-join queues and split-merge queues that are under the same job arrival process and with the same sub-queue's service time distribution are called $(\lambda,\mu)$-equivalent queues to each other, where $\lambda$ and $\mu$ are the job arrival rate and the sub-queue's service rate respectively.}
\end{definition}

\subsection{Approximations for General Fork-Join Queues}
This paper uses the term \texttt{general queues} to denote the fork-join queues with identically and generally distributed sub-queues' service times and job inter-arrival times.
\begin{theorem}[LT of Sojourn Time]{
The sojourn time of a general non-purging $(n,k)$ fork-join queue can be represented by a linear combination of the sojourn times of the $(\lambda,\mu)$-equivalent basic fork-join queues:
\begin{equation}
t_{(n,k)}=\sum_{i=k}^{n}W_i^{n,k}t_{(i,i)}
\end{equation}
 where $W_i^{n,k}$ is the corresponding $W$ coefficient.
}\label{the5}
\begin{proof}
The only difference between the $(\lambda,\mu)$-equivalent non-purging $(n,k)$ queue and basic $(n,n)$ queue lies in the job departure process, which has no influence over the job arrival process, the sub-queue's service time distribution and therefore the distribution of the sub-queue length. Consequently, the target non-purging $(n,k)$ queue and its $(\lambda,\mu)$-equivalent $(n,n)$ fork-join queue have the same family of jointly-identical rvs  $t_{1\isep n}$, and therefore the same order statistic $t_{(n,k)}$.

According to Lemma \ref{the4} and Theorem \ref{the2}, $t_{(n,k)}=\sum_{i=k}^{n}W_i^{n,k}t^{sub}_{(i,i)}$, where $t^{sub}_{(i,i)}$ is the maximum stable sojourn time of $i$ arbitrarily chosen sub-queues from the target non-purging $(n,k)$ queue. By the constitution methodology of fork-join queues, the $i$ chosen sub-queues can constitute an $(i,i)$ queue which is $(\lambda,\mu)$-equivalent to the target non-purging $(n,k)$ queue. Therefore, $t_{(i,i)}=t^{sub}_{(i,i)}$ and $t_{(n,k)}=\sum_{i=k}^{n}W_i^{n,k}t^{sub}_{(i,i)}=\sum_{i=k}^{n}W_i^{n,k}t_{(i,i)}$ hold.
\end{proof}
\end{theorem}

\begin{theorem}[LT of Expected Sojourn Time]{The expected sojourn time $\mathbb {NT}_{n,k}\equiv E[t_{(n,k)}]$ of a general non-purging $(n,k)$ fork-join queue can be represented by a linear combination of the expected sojourn times of the $(\lambda,\mu)$-equivalent basic fork-join queues:
\begin{equation}
\mathbb {NT}_{n,k}=\sum_{i=k}^{n}{W_i^{n,k}\mathbb T_{i}}
\end{equation}
where $\mathbb T_{i}$ is the expected sojourn time of the $(\lambda,\mu)$-equivalent basic $(i,i)$ fork-join queue and $W_i^{n,k}$ is the corresponding $W$ coefficient.
}\label{the6}
\begin{proof}
$
E[t_{(n,k)}]=E[\sum_{i=k}^{n}W_i^{n,k}t_{(i,i)}]=\sum_{i=k}^{n}W_i^{n,k}\mathbb T_{i}
$
\end{proof}

\end{theorem}

\begin{remark}
The independence of the service times of the sub-queues is not required for Theorem \ref{the5} and \ref{the6} to hold. To put Theorem \ref{the6} into practices, we need to find the existing computing methods of $\mathbb T_{i}$.
\end{remark}

\subsection{Approximations for Exponential Fork-Join Queues}

This paper uses the term \texttt{exponential queues} to denote the fork-join queues under a Poisson job arrival process and with iid exponential sub-queues' service time distributions.

For exponential $(n,n)$ fork-join queues, there exist two reliable approximation methods: Nelson's approximation\cite{Nelson:1988jk} and Varma's approximation\cite{Varma:1994gm}. Accordingly, we can propose two approximation methods for exponential non-purging $(n,k)$ fork-join queues: The \texttt{Nelson-LT approximation} based on Nelson's approximation and the \texttt{Varma-LT approximation} based Varma's approximation.

\subsubsection{The Nelson-LT Approximation}
\begin{figure*}[ht!]
\centering
\includegraphics{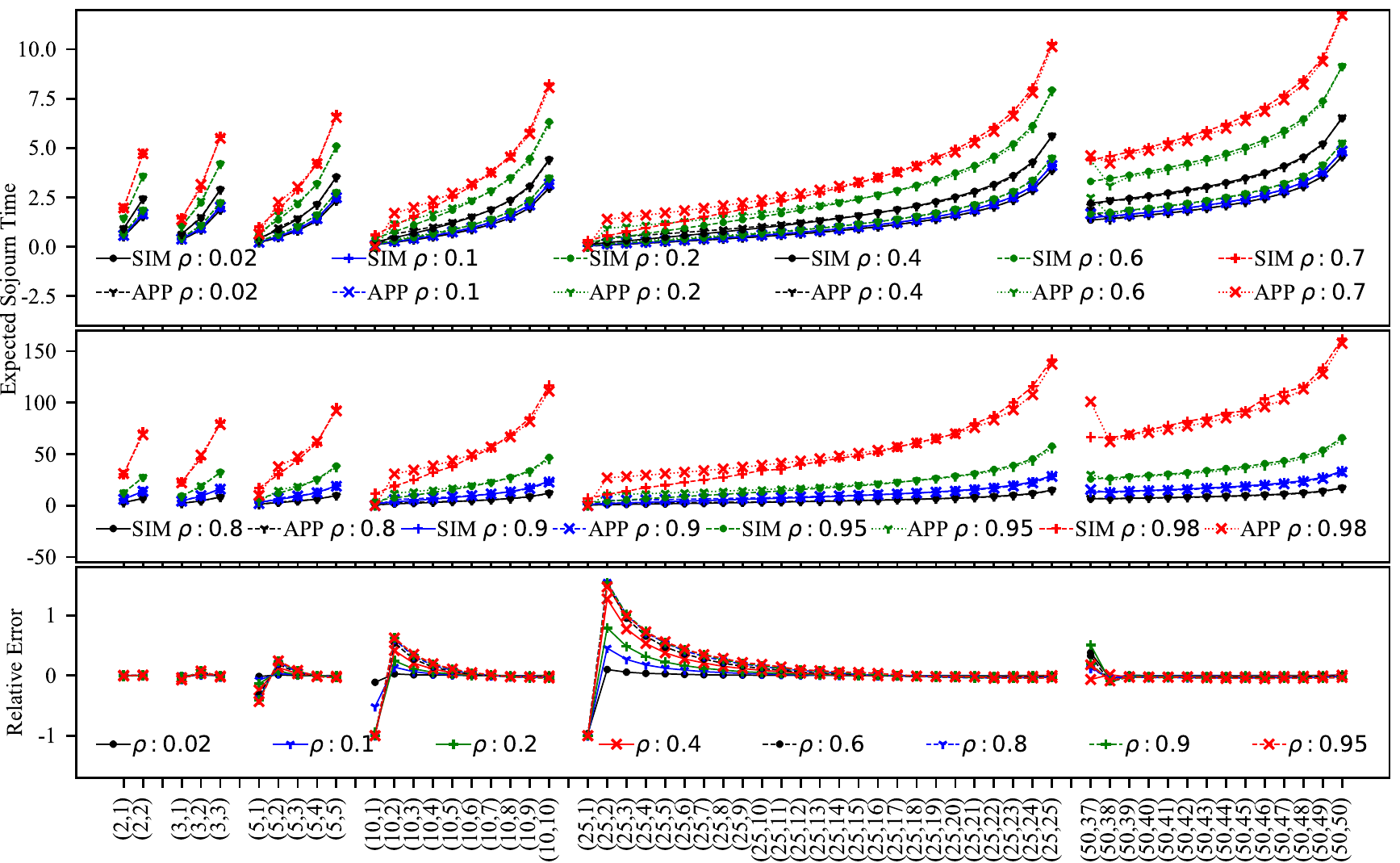}
\caption{The Nelson-LT Approximations for Exponential Non-Purging $(n,k)$ Fork-Join Queues ($\mu=1$, SIM: Simulation, APP: approximation)}
\label{linear-nelson}
\end{figure*}
For exponential $(n,n)$ fork-join queues, Nelson et al. \cite{Nelson:1988jk} proposed the following approximations:
\begin{gather}
\mathbb T_1=\frac{1}{\mu(1-\rho)}, \quad\quad \mathbb T_2=\frac{12-\rho}{8}\mathbb T_1\notag \\
\mathbb T_n\simeq\left[\frac{H_n}{H_2}+\frac{4}{11}\left(1-\frac{H_n}{H_2}\right)\rho\right]\mathbb T_2, n\geq 2,\notag
\end{gather}
where $\lambda$ and $\mu$ are respectively the job arrival rate and the sub-queue's service rate, $\rho\equiv \frac{\lambda}{\mu}$ is called the load factor of the queue, $\mathbb T_n$ is the expected sojourn time of $(n,n)$ basic fork-join queue, and $H_n=\sum_{i=1}^n\frac{1}{i}$ is called the harmonic number. Consequently, our approximations can be specified in the following theorem.
\begin{theorem}[Nelson-LT Approximation]{According to Nelson's approximation, the expected sojourn time $\mathbb {NT}_{n,k}$ of an exponential non-purging $(n,k)$ fork-join queue can be approximated as follow:
\begin{align}\label{exp-app}
\mathbb {NT}_{n,k}\simeq \left\{
\begin{array}{ll}
\begin{array}{c}
\frac{n}{\mu(1-\rho)}+\frac{12-\rho}{88\mu(1-\rho)}\times \\
\sum_{i=2}^{n}{W_i^{n,1}\left[\frac{11H_i+4\rho(H_2-H_i)}{H_2}\right]}\end{array}&k=1,\\
\frac{12-\rho}{88\mu(1-\rho)}\sum_{i=k}^{n}{W_i^{n,k}\left[\frac{11H_i+4\rho(H_2-H_i)}{H_2}\right]}&k\geq 2.
\end{array}
\right.
\end{align}
  where $\lambda$ and $\mu$ are respectively the job arrival rate and the sub-queue's service rate, and $\rho\equiv \frac{\lambda}{\mu}$ is the load factor of the target queue. Specially, we replace any negatively approximated $\mathbb {NT}_{n,k}$ with 0.
}\label{non-e}
\end{theorem}

We examine above linear-transformation approximations against the mean sojourn times of jobs sampled from various simulated exponential non-purging fork-join queues (details of simulations in this paper can be found in Appendix). The value of $W$ coefficients used in Eq. \ref{exp-app} are given by Eq. \ref{eq-w-coe}. The results depicted in Fig. \ref{linear-nelson} confirmed the validity of our technique under a moderate value of $n$ ($n\leq 50$) and a relatively large value of $k$ (compared to $n$). The relative errors are calculated as $\frac{APP}{SIM}-1$.

We notice that when $k$ is relatively small, the approximation tends to be uncontrollable, which can be due to the fact that the smaller $k$ is, the more items in Eq. \ref{exp-app} are summed, and consequently, the more relative errors introduced by Nelson's approximations are accumulated.

These results also confirmed the high-precision merit of Nelson's approximations, since $W$ coefficients tend to be very large with the increase of $n$, for example $W^{25,9}_{16}=13146544125$. As a result, the relative error introduced by Nelson's approximation has to be amplified by the large value of the corresponding $W$ coefficient.

\subsubsection{The Varma-LT Approximation}
\begin{figure*}[ht!]
\centering
\includegraphics[width=\textwidth]{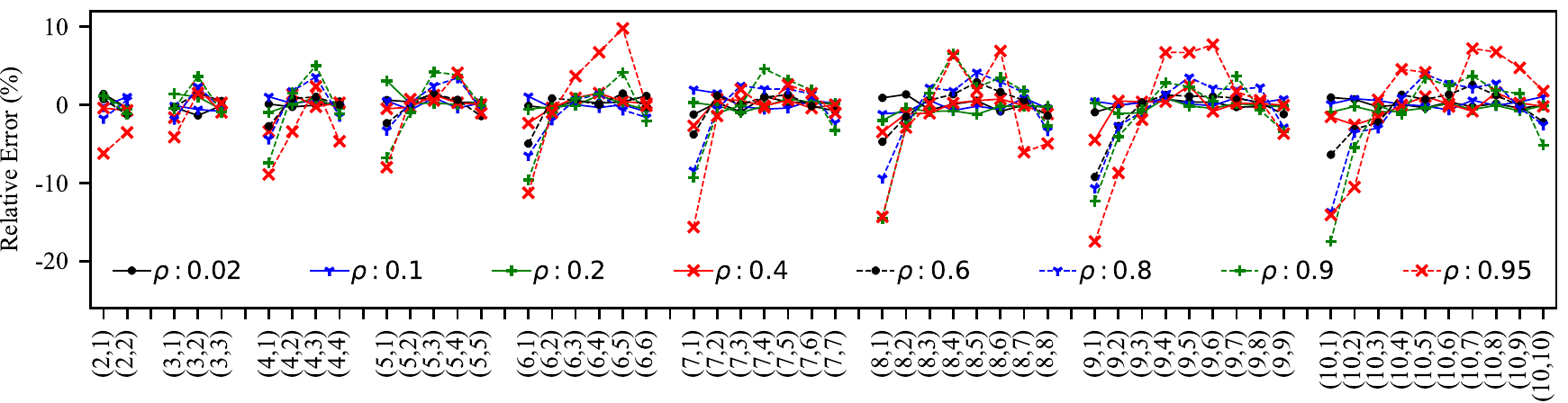}
\caption{The Relative Errors of the Varma-LT Approximations for Exponential Non-Purging $(n,k)$ Fork-Join Queues}
\label{linear-varma}
\end{figure*}

For exponential non-purging $(n,n)$ fork-join queues, Varma et al. \cite{Varma:1994gm} gave another well-known approximation method based on the so-called light traffic interpolation technique. The expected sojourn time is approximated as
$$
\mathbb T_n\simeq\left[H_n+(V_n-H_n)\frac{\lambda}{\mu}\right]\frac{1}{\lambda-\mu}, 0\leq \lambda<\mu
$$
where $V_n= \sum_{r=1}^{n} {n \choose r}(-1)^{r-1}\sum_{m=1}^{r}{r\choose m}\frac{(m-1)!}{r^{m+1}}$.
 
As our linear transformation technique is orthogonal to the concrete approximation methods for basic fork-join queues, we replace Nelson's approximation with the above Varma's approximation to try to avoid the approximations' uncontrollability appeared in Theorem \ref{non-e}. Consequently, the new approximations can be specified in the following theorem.
\begin{theorem}[Varma-LT Approximation]{According to Varma's approximation, the expected sojourn time of an exponential non-purging $(n,k)$ fork-join queue can be approximated as follow:
\begin{equation}\label{eq-lt-varma}
\mathbb {NT}_{n,k}\simeq\sum_{i=k}^{n}{W_i^{n,k}\left[H_i+(V_i-H_i)\frac{\lambda}{\mu}\right]\frac{1}{\lambda-\mu}}
\end{equation}
  where $\lambda$ and $\mu$ are respectively the job arrival rate and the sub-queue's service rate.
}\label{non-e-varma}
\end{theorem}
We examine Eq. \ref{eq-lt-varma} against the mean sojourn times of jobs sampled from various simulated non-purging fork-join queues. The employed $W$ coefficients are calculated from Eq. \ref{eq-w-coe}. The results depicted in Fig. \ref{linear-varma} showed that the new approximations are fairly good when $n\leq 10$ and $\rho$ is not too extreme ($\rho\leq 0.9$): The relative error is generally less than 10\%, which is much more controllable than the approximations given by Theorem \ref{non-e}.

However, as the Varma's approximation itself becomes out of control when $n\geq 55$, Theorem \ref{non-e} is more valuable in general cases.

\section{Bounds for Purging $(n,k)$ Fork-Join Queues}\label{bounds}
Unlike in non-purging $(n,k)$ fork-join queues, the sojourn time distribution of a sub-task in purging $(n,k)$ fork-join queues changes when either $n$ or $k$ varies, and thus differs from the sojourn time distribution of a sub-task in the $(\lambda,\mu)$-equivalent basic fork-join queues. As a result, we cannot build similar linear-transformation approximations for purging queues. However, the expected sojourn times of non-purging queues can serve as the upper bounds of the $(\lambda,\mu)$-equivalent purging queues' expected sojourn times.

\subsection{The Naive Upper Bounds}
\begin{theorem}[Naive Upper Bounds]{The expected sojourn time $\mathbb {PT}_{n,k}$ of a purging $(n,k)$ fork-join queue can be upper bounded as follow:
\begin{equation}
  \mathbb {PT}_{n,k}\leq \sum_{i=k}^{n}{W_i^{n,k}\mathbb T_{i}}\label{uppeq}
\end{equation}
}\label{Upper-bounds}
 where $\mathbb T_{i}$ is the expected sojourn time of the $(\lambda,\mu)$-equivalent basic fork-join queue.
\begin{proof}
The right side of Eq. \ref{uppeq} is the expected sojourn time of the $(\lambda,\mu)$-equivalent non-purging $(n,k)$ fork-join queue. As the expected sub-queue length of a stable purging $(n,k)$ fork-join queue is no longer than that of the $(\lambda,\mu)$-equivalent stable non-purging queue, the expected sojourn time of the purging $(n,k)$ fork-join queue is thus no larger than that of its non-purging $(\lambda,\mu)$-equivalent queue.
\end{proof}
\end{theorem}

\paragraph{Comparing with Existing Stat-of-the-Art Upper Bounds}
For purging $(n,k)$ fork-join queues with iid service time distributions and under a Poisson job arrival process, Existing state-of-the-art upper bounds on the expected sojourn time are the \textit{split-merge upper bounds} given by Joshi et al. \cite{Joshi:2017bj}:
\begin{equation}\label{josh-upper-bound}
  \mathbb {PT}_{n,k}\leq E[X_{(n,k)}]+ \frac{\lambda E[X_{(n,k)}^2]}{2(1-\lambda E[X_{(n,k)}])}
\end{equation}
 where $\lambda$ is the job arrival rate and $X_{(n,k)}$ is the $k^{th}$ order statistic of the iid service time rvs $X_{1\isep n}$.
 
The right side of Eq. \ref{josh-upper-bound} is the expected sojourn time of the $(\lambda,\mu)$-equivalent $(n,k)$ split-merge queue.

\begin{corollary}{Split-merge upper bounds become much looser than naive upper bounds when $E[X_{(1,1)}]< E[X_{(n,k)}]$ and $\lambda \rightarrow [\frac{1}{E[X_{(n,k)}]}]^-$.}\label{torhocoro}
\begin{proof}
When $\lambda \rightarrow [\frac{1}{E[X_{(n,k)}]}]^-$, the bounds given by Eq. \ref{josh-upper-bound} approach $+\infty$. Moreover, the bounds become meaningless when $\lambda \geq \frac{1}{E[X_{(n,k)}]}$. As a contrast, our naive bounds are finite meaningful values, as long as  $\lambda < \left[\mu\equiv\frac{1}{E[X_{(1,1)}]}\right]$. Besides, $\mu>\frac{1}{E[X_{(n,k)}]}$.
\end{proof}
\begin{remark}
Apparently there is a range of load factors $\rho \in [\frac{E[X_{(1,1)}]}{E[X_{(n,k)}]},1)$ cannot be bounded by Eq. \ref{josh-upper-bound}, and the larger $k$ is, the smaller bounded-able $\rho$ range becomes, while the naive bounds are applicable as long as $\rho<1$.
\end{remark}
\end{corollary}
\begin{corollary}{Naive upper bounds become much looser than split-merge upper bounds when $k \rightarrow 1$.}\label{to1coro}
\begin{proof}
When $k \rightarrow 1$, more and more sub-tasks are purged from both the sub-queues and the service stations when the $k^{th}$ finished sub-task is acknowledged by the purging $(n,k)$ queue, as a result of which, the expected sub-queue length becomes shorter and shorter than that of the $(\lambda,\mu)$-equivalent non-purging $(n,k)$ queue. On the contrary, the expected sub-queue length of the target purging $(n,k)$ fork-join queue becomes closer and closer to that of the $(\lambda,\mu)$-equivalent $(n,k)$ split-merge queue. At last, the purging $(n,1)$ fork-join queue equates to the $(n,1)$ split-merge queue, which gives us the following exact closed-form solution:
$\mathbb {PT}_{n,1}= E[X_{(n,1)}]+ \frac{\lambda E[X_{(n,1)}^2]}{2(1-\lambda E[X_{(n,1)}])}$.
\end{proof}
\begin{remark} On the other side of Corollary \ref{to1coro}, when $k \rightarrow n$, the expected sub-queue length of the purging $(n,k)$ fork-join queue becomes closer and closer to that of the $(\lambda,\mu)$-equivalent non-purging queues. At last, the purging $(n,n)$ fork-join queue equates to the $(\lambda,\mu)$-equivalent non-purging $(n,n)$ fork-join queue.
\end{remark}
\end{corollary}

\subsection{The Refined Upper Bounds}

\begin{figure*}[ht!]
\centering
\subfloat[Naive Upper Bounds (Naive) v.s. Split-Merge Upper Bounds (SM)]{\includegraphics{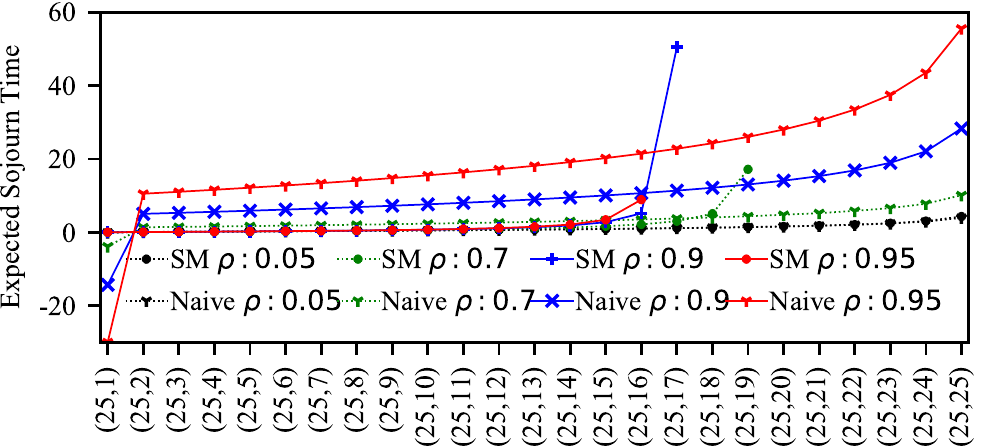}}
\hfill
\subfloat[Refined Bounds (RF) v.s. Simulations (SIM)]{\includegraphics{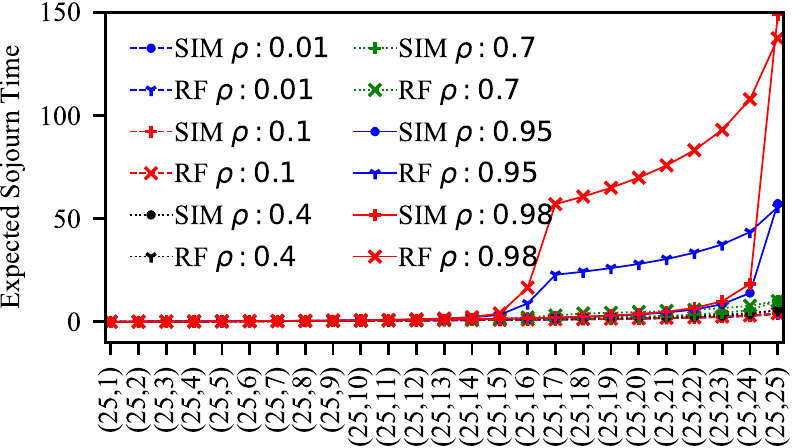}}
\caption{Upper Bounds for Exponential Purging $(25,1),(25,2),...,(25,25)$ Fork-Join Queues ($\mu=1$)}
\label{Up-fig}
\end{figure*}

\begin{theorem}[Refined Upper Bounds]{For a purging $(n,k)$ fork-join queue with iid service time distributions and under a Poisson job arrival process, the expected sojourn time $\mathbb {PT}_{n,k}$ can be upper bounded as follows:
\begin{align}
  \mathbb {PT}_{n,k}\!\!\leq\!\!\left\{
\begin{array}{cl}\label{eq-r}
\sum_{i=k}^{n}{W_i^{n,k}\mathbb T_{i}}&\!\!\!\!\!\!\!\!\!\!\!\lambda \geq \frac{1}{E[X_{(n,k)}]},\\
\!\!\!\!\min\!\left(\sum_{i=k}^{n}\!\!W_i^{n,k}\mathbb T_{i},\begin{array}{c}E[X_{(n,k)}]\!+\\
\!\frac{\lambda E[X_{(n,k)}^2]}{2(1-\lambda E[X_{(n,k)}])}\end{array}\!\right)& \!\!\!otherwise.
\end{array}
\right.
\end{align}
}\label{r-Upper-bounds}
 where $\mathbb T_{i}$ is the expected sojourn time of the $(\lambda,\mu)$-equivalent $(i,i)$ fork-join queue, $\lambda$ is the job arrival rate, and $X_{(n,k)}$ is the $k^{th}$ order statistic of the iid service time rvs $X_{1\isep n}$.
\begin{proof}
According to Corollary \ref{torhocoro} and \ref{to1coro}, and excluding the split-merge bounds when $\lambda \geq \frac{1}{E[X_{(n,k)}]}$, we derive Eq. \ref{eq-r}.
\end{proof}
\begin{remark} Although Eq. \ref{eq-r} has extended the bounded-able range of $\rho$ from $[0,\frac{E[X_{(1,1)}]}{E[X_{(n,k)}]})$ to $[0,1)$, there is still an untamed range of $\rho$, since purging $(n,k)$ fork-join queues may still keep stable even when $\rho\geq 1$.
\end{remark}
\end{theorem}

\paragraph{Upper Bounds for Exponential Queues}
Specially, we give the refined upper bounds for exponential purging $(n,k)$ fork-join queues.
\begin{theorem}[Refined Upper Bounds for Exponential Purging Queues]{For exponential purging $(n,k)$ fork-join queues, the expected sojourn time $\mathbb {PT}_{n,k}$ can be upper bounded as follows:
\begin{align}\label{eq-eu}
  \mathbb {PT}_{n,k}\!\!\leq\!\!\left\{
\begin{array}{c}
\underbrace{\!\!\!\!\!\begin{array}{c}\frac{12-\rho}{88\mu(1-\rho)}	
\sum_{i=k}^{n}{W_i^{n,k}\!\left[\frac{11H_i+4\rho(H_2-H_i)}{H_2}\right]}\end{array}}_{Naive}\\
\text{when }k\geq 2 \text{ and }\rho \geq \frac{1}{H_n-H_{n-k}},\\\\
\!\!\!\!\min\!\left(Naive,\overbrace{\begin{array}{c}
	\frac{H_n-H_{n-k}}{\mu}+\\
\frac{\rho[(H_{n^2}-H_{(n-k)^2})+(H_n-H_{(n-k)})^2]}{2\mu[1-\rho(H_n-H_{n-k})]}
	\end{array}}^{Split-Merge}\right)\\
	otherwise.
\end{array}
\right.
\end{align}
}\label{e-Upper-bounds}  
 where $\lambda$ and $\mu$ are respectively the job arrival rate and the sub-queue's service rate, $\rho\equiv \frac{\lambda}{\mu}$, and $H_{n^2}=\sum^{n}_{i=1}\frac{1}{i^2}$.
 
\begin{proof}
The split-merge part of the Eq. \ref{eq-eu} is already given by Theorem 2 of \cite{Joshi:2012cv}. According to Theorem \ref{non-e} and \ref{r-Upper-bounds}, we derive Eq. \ref{eq-eu}.
\end{proof}
\end{theorem}
We make numerical comparisons between naive upper bounds and split-merge upper bounds, and examine refined upper bounds against the mean sojourn times of jobs sampled from various simulated purging fork-join queues. The value of $W$ coefficients used in Eq. \ref{eq-eu} are given by Eq. \ref{eq-w-coe}. We find that:
\begin{itemize}
  \item The split-merge upper bounds become extremely pessimistic with the increase of $k$, but tend to be much tighter than naive bounds when $k$ is small (see Fig. \ref{Up-fig} (a)). These results are consistent with Corollary \ref{torhocoro} and \ref{to1coro}.
  \item There is still plenty of room of improving the upper bounds when k is relatively large (see Fig. \ref{Up-fig} (b)).
\end{itemize}
 
\subsection{Lower Bounds}
To complete our work, we review and compare the state-of-the-art lower bounds for purging $(n,k)$ fork-join queues.

For purging $(n,k)$ fork-join queues with iid service time distributions and under a Poisson job arrival process,
the state-of-the-art lower bounds are the \textit{split-merge lower bounds} given in \cite{Joshi:2017bj}:
\begin{equation}\label{eq-sm-b}
 \mathbb {PT}_{n,k}\geq E[X_{(n,k)}]+\frac{\lambda E[X_{(n,1)}^2]}{2(1-\lambda E[X_{(n,1)}])}
\end{equation}
 where $\lambda$ is the job arrival rate and $X_{(n,k)}$ is the $k^{th}$ order statistic of the iid service time rvs $X_{1\isep n}$.

For exponential purging $(n,k)$ fork-join queues, there is another staging analysis based lower bound \cite{Joshi:2012cv}:
\begin{equation}
\mathbb {PT}_{n,k}\geq\frac{H_n-H_{n-k}}{\mu}+\frac{\rho(H_{n(n-\rho)}-H_{(n-k)(n-k-\rho)})}{\mu}
\end{equation}
where $\lambda$ and $\mu$ are respectively the job arrival rate and the sub-queue's service rate, $\rho\equiv \frac{\lambda}{\mu}$, and $H_{n(n-\rho)}=\sum_{i=1}^{n}\frac{1}{i(i-\rho)}$. 

This \textit{staging lower bound} is adapted from the staging lower bound for basic fork-join queues proposed in \cite{Varki:2001wc}, which requires a memory-less property of the service time distribution. Accordingly, this bound cannot be applied to general purging queues.

\begin{theorem}{\label{th-better-lower}
The staging lower bounds are tighter than the split-merge lower bounds.}
\begin{proof}
The exact form of Eq. \ref{eq-sm-b} for exponential queues can be transformed into: $$\mathbb {PT}_{n,k}\geq\frac{H_n-H_{n-k}}{\mu}+\frac{\rho (H_{n(n-\rho)}-H_{(n-1)(n-1-\rho)})}{\mu}.$$ As $H_{(n-k)(n-k-\rho)})< H_{(n-1)(n-1-\rho)})$ when $k>1$, we derive Theorem \ref{th-better-lower}.
\end{proof}
\end{theorem}
\begin{figure}[tp!]
\centering
\includegraphics{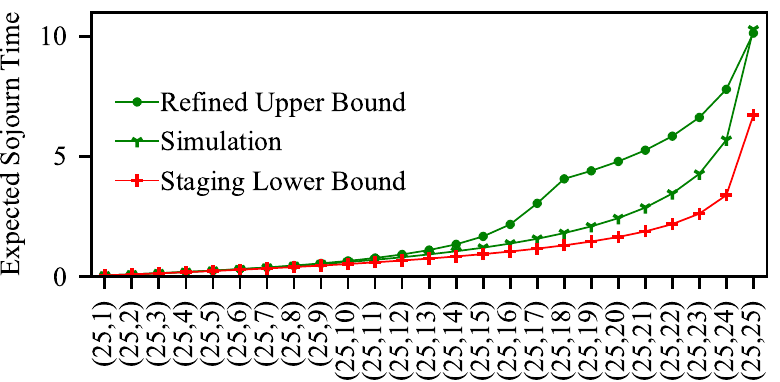}
\caption{Bounds v.s. Simulations for Exponential Purging $(25,1), (25,2),..., (25,25)$ Fork-Join Queues ($\mu=1, \rho=0.7$)}
\label{lower-fig}
\end{figure}
 We examine bounds for exponential purging queues against simulations. Fig. \ref{lower-fig} depicts the large gap between the upper bounds and the lower bounds when $k$ is relatively large, due to which, we can hardly find reasonable approximations of the expected sojourn time of purging $(n,k)$ fork-join queues.
 
\section{Discussion}\label{discuss}

\begin{figure}[tp!]
\centering
\includegraphics{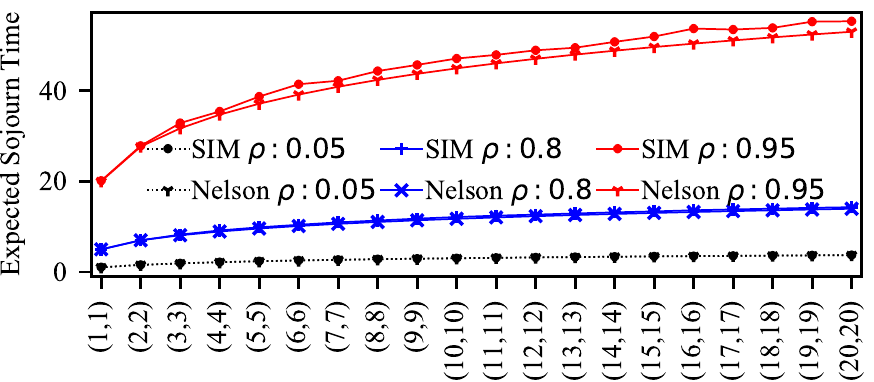}
\caption{Simulations (SIM) v.s. Nelson's Approximations for Basic $(1,1),(2,2),...,(20,20)$ Fork-Join Queues ($\mu$=1)} \label{simu-nelson}
\end{figure}

Currently, there is an unnegligible limitation when put the new proposed linear transformation technique into practices: The value of $W$ coefficient given by Eq. \ref{eq-w-coe} increases explosively with the increase of n, for example $W^{40,37}_{37}={40 \choose 37}=9880$, $W^{50,37}_{37}={50 \choose 37}=354860518600$, and $W^{100,37}_{37}={100 \choose 37}=3.42002954749393 \times 10^{27}$, as a result of which, the original negligible relative error of $\mathbb T_i$ in Theorem \ref{the6} will be amplified into a huge deviation of the approximated $\mathbb {NT}_{n,k}$. Consequently, the practicability of the linear transformation technique depends on whether we can find high-precision approximated or simulated $\mathbb T_i$. For example, when we use simulated $\mathbb T_i$ to estimate $\mathbb {NT}_{n,k}$, the results are far from acceptable (see Table \ref{simu-based}), and also far behind the Nelson-LT approximations (see Fig. \ref{linear-nelson}). These surprising results can be due to the fluctuation of the simulated $\mathbb T_i$ (see Fig. \ref{simu-nelson}). Fig. \ref{linear-nelson} has depicted that the accuracy of the Nelson-LT approximation is acceptable when $n\leq 50$ and $k$ is relatively large (for example, $k> 37$ when $n=50$). However the approximations are similarly unacceptable when $k$ is relatively small (see Table \ref{nel-based}).
\begin{table}[h!]
\centering
\caption{Approximated $\mathbb {NT}_{n,k}$ Based on Simulated $\mathbb T_i$ ($\mu=1$)}
\label{simu-based}
\begin{tabular}{l|lllll}
$(n,k)$   & $\rho$: 0.05   & $\rho$ :0.4    & $\rho$: 0.8\\
\hline
(20,17) & 25.17029838  & -37.93974618 & 70.76161205\\
(20,18) & 0.412468275  & 6.238904164  & 3.024177347\\
(20,19) & 2.772567927  & 3.796009747  & 11.47066833
\end{tabular}
\end{table}
\begin{table}[h!]
\centering
\caption{Approximated $\mathbb {NT}_{n,k}$ Based on Nelson's $\mathbb T_i$ ($\mu=1$)}
\label{nel-based}
\begin{tabular}{l|lllll}
$(n,k)$   & $\rho$: 0.2   & $\rho$ :0.4    & $\rho$: 0.8\\
\hline
(50,34) & -317.7265625 & 	-203.25  & -580.0859375\\
(50,35) & 23	 & -7.599609375	 & -18.87890625\\
(50,36) & -4.269042969 & 	4.208007813  & 9.935546875
\end{tabular}
\end{table}

The fundamental solution of this problem is to scale down $W$ coefficients,  a promising research direction needs further efforts.

 Anyway, this linear transformation technique is capable of estimating/bounding the performance of most of practical non-purging/purging $(n,k)$ fork-join queueing systems, where the replication factor ($n$) rarely exceeds 10, a result of cost-effective tradeoff. For example, the replication factor of either Dynamo or Cassandra is commonly 3. Under such configurations, a write operation in Dynamo/Cassandra will be forked into 3 copies exactly. For such $n\leq 10$ cases, we have proposed the fairly good Varma-LT approximations (Theorem \ref{non-e-varma}).

From another perspective, the linear transformation technique can be used to check simulators' precision and to find better closed-form approximations for basic fork-join queues.

\section{Related Works}\label{review}
\paragraph{Order Statistics}
Bertsimas et al. \cite{Bertsimas:2006ge} gave some tight bounds on the expectation of the $k^{th}$ order statistic given the first and second moment information on $n$ real-valued rvs. We gave exact linear transformations for $k^{th}$ order statistic instead of bounds. Shi et al. \cite{Shi:2013gw} proposed a dynamic programming algorithm to compute the order statistics of a family of correlated rvs whose distributions are not required to be identical. This algorithm relies on the existence of computing methods for the distributions of both minimum and maximum of target rvs. As a contrast, our work is more formal and easier to practice for the reveal of the closed-form linear transformation, and only relies on the existence of computing methods for the maximum's distribution.
  
\paragraph{Basic Fork-Join Queues}
For $n=2$ fork-join queues under a Poisson job arrival process: 1) Flatto et al. \cite{Flatto1984} gave the queue length distribution for exponential queues in stable state; 2) Baccelli \cite{baccelli1985two} extended Flatto's work to queues with general service time distributions; 3) Nelson et al. \cite{Nelson:1988jk} proposed the exact closed-form solution of the expected sojourn time for exponential queues in stable state.

For $n\geq 3$ exponential fork-join queues, the most influential approximation work was proposed by Nelson et al. in 1988 \cite{Nelson:1988jk}, which is based on the fact that the sojourn times $X_{1\isep k}$ of sub-tasks $1, 2,...,k$ are associated rvs \cite{Esary:1967eo}, whose maximum can be bounded by the maximum of their iid equivalents. The lower bound is obtained by neglecting queueing effects. The approximation is a linear mixture of the upper and lower bounds. Parameters of the mixture are learned based on the mean sojourn times of jobs sampled from simulated basic fork-join queues.
Varki et al. \cite{Varki:2001wc} improved the lower bound by using an staging analysis technique \cite{Trivedi:1982vp} based on the memory-less property of the exponential service time distribution, and use the mean value of Nelson's upper bound and the staging lower bound as the approximation. According to experiments in \cite{Lebrecht:2007tm}, Nelson's approximation is still the most reliable one for exponential queues, compared to following works including \cite{Varma:1994gm} and \cite{Varki:2001wc}.

Varma et al. \cite{Varma:1994gm} extended Nelson's approximation to general service time distributions using a light traffic interpolation technique.
Thomasian et al. \cite{Thomasian:1994kq} employed linear regression over the statistics of simulated fork-join jobs to find the parameters of their approximation equation for the expected sojourn time. However any change in service time distributions will require for re-simulations and re-regressions.
Recently, Rizk et al. \cite{Rizk:2015tn} proposed the first computable bounds on waiting and sojourn time of fork-join queues with general service time distributions by using martingales. However the upper bound is looser than Nelson's when it comes to the exponential service time distribution. Fidler et al. \cite{Fidler:2016tw} considered the multi-stage nature of many fork-join queue networks, and proposed their end-to-end delay bounds.

We refer readers to \cite{Thomasian:2014df} for a more comprehensive survey on fork-join queuing systems. To conclude, our work is orthogonal to existing approximation methods for basic fork-join queues.

\paragraph{Purging $(n,k)$ Fork-Join Queues}

There are some exact quantity analyses \cite{Gardner:2015kb,Lee:2017gi} for purging $(n,1)$ fork-join queues, as it is equivalent to a single queue with $n$ times the service rate. Gardner et al. \cite{Gardner:2015kb} gave comprehensive research on purging $(n,1)$ fork-join queues, with considerations on multi-class jobs, interferences from un-forked jobs and heterogeneous service time distributions. Lee et al. \cite{Lee:2017gi} take the purging overheads into consideration, since the cancellation of running jobs typically incurs unnegligible delays in practice.

For purging $(n,k>1)$ fork-join queues, there are even no applicable approximations currently. Joshi et al. \cite{Joshi:2012cv} extended the staging analysis to exponential $(n,k)$ fork-join queues to find the lower bounds. Bounds for queues with general service time distributions are given by \cite{Joshi:2012cv} and \cite{Joshi:2017bj}, by resorting the fork-join queue to the split-merge queue model, where all empty sub-queues are blocked until any $k$ sub-tasks of the current job are completed. As depicted in Fig. \ref{Up-fig} (a), the proposed upper bounds tend to be very rough when increasing $k$ or the load factor $\rho$.

\paragraph{Non-Purging $(n,k)$ Fork-Join Queues}
A typical use case of non-purging $(n,k)$ fork-join queues is the writing process in Cassandra \cite{Huang:2014gq}.
Fidler et al. \cite{Fidler:2016tw} gave non-asymptotic statistical bounds on the sojourn time of non-purging $(n,k)$ fork-join queues. As a contrast, we give proper approximations instead of bounds.

\section{Conclusion and Future Work}\label{con}

Despite the popularity of $(n,k)$ fork-join queues, there were no practical existing approximations of their expected sojourn times. Only some rough bounds have been given, which tend to be extremely loose when increasing $k$ or the load factor $\rho$. This paper gave the first applicable approximation method for non-purging $(n,k)$ fork-join queues and tackled the uncontrollability of the bounds for purging $(n,k)$ fork-join queues:
\begin{itemize}
  \item A brand-new closed-form linear transformation technique is developed for jointly-identical rvs, which provides a bridge to reduce the sojourn time approximation problem of non-purging $(n,k)$ fork-join queues to that of basic fork-join queues.
  \item Improvements over upper bounds on the expected sojourn time of purging $(n,k)$ fork-join queues are also gained by resorting the purging queues to their non-purging $(\lambda,\mu)$-equivalents.
\end{itemize}

Above innovations are examined by simulation experiments and numerically compared to the stat-of-the-arts. Results show that this linear transformation approach is practiced well for exponential $(n,k)$ fork-join queues with moderate $n$ and relatively large $k$. However, as currently found $W$ coefficients (coefficients of the linear combination) increase explosively with the increase of $n$, there is an uncontrollable deviation in new proposed approximations when $n$ is large and $k$ is relatively small. Fortunately, approximations for real-life fork-join systems are unlikely influenced by this problem.

In the future, more efforts should be put into: Scaling down $W$ coefficients, improving the approximations for basic fork-join queues with the help of the linear transformation technique, and evaluating the performance of real-life $(n,k)$ fork-join systems as complement to existing experimental methods \cite{Wang:2014tr,Kuhlenkamp:2014te}.

\appendix
The simulator employed by this work is Forkulator-p: https://github.com/excelwang/forkulator-p, which is modified from Forkulator \cite{Fidler:2016wa} with additional features of simulating purging queues. For each $(n,k)$ pair, the sub-queue's service rate $\mu$ is set to 1, simulated jobs are sampled at a rate of 1\%, and the mean sojourn times are calculated on 10000 samples.

The solver for currently found $W$ coefficients and some pre-calculated values can be accessed from: https://github.com/excelwang/WCoefficients. We give some frequently used $W^{n,k}_i$ coefficients bellow.\\\\
{\centering
{\scriptsize
$\arraycolsep=2.5pt\def\arraystretch{1.8533405}
\begin{array}{lllll}
\hline
W^{3,1}_1: 3	&	W^{6,3}_4: -45	&	W^{8,1}_7: 8	&	W^{9,2}_6: 420	&	W^{10,2}_3: -240	\\
W^{3,1}_2: -3	&	W^{6,3}_5: 36	&	W^{8,1}_8: -1	&	W^{9,2}_7: -216	&	W^{10,2}_4: 630	\\
W^{3,1}_3: 1	&	W^{6,3}_6: -10	&	W^{8,2}_2: 28	&	W^{9,2}_8: 63	&	W^{10,2}_5: -1008	\\
W^{3,2}_2: 3	&	W^{6,4}_4: 15	&	W^{8,2}_3: -112	&	W^{9,2}_9: -8	&	W^{10,2}_6: 1050	\\
W^{3,2}_3: -2	&	W^{6,4}_5: -24	&	W^{8,2}_4: 210	&	W^{9,3}_3: 84	&	W^{10,2}_7: -720	\\
W^{3,3}_3: 1	&	W^{6,4}_6: 10	&	W^{8,2}_5: -224	&	W^{9,3}_4: -378	&	W^{10,2}_8: 315	\\
W^{4,1}_1: 4	&	W^{6,5}_5: 6	&	W^{8,2}_6: 140	&	W^{9,3}_5: 756	&	W^{10,2}_9: -80	\\
W^{4,1}_2: -6	&	W^{6,5}_6: -5	&	W^{8,2}_7: -48	&	W^{9,3}_6: -840	&	W^{10,2}_{10}: 9	\\
W^{4,1}_3: 4	&	W^{6,6}_6: 1	&	W^{8,2}_8: 7	&	W^{9,3}_7: 540	&	W^{10,3}_3: 120	\\
W^{4,1}_4: -1	&	W^{7,1}_1: 7	&	W^{8,3}_3: 56	&	W^{9,3}_8: -189	&	W^{10,3}_4: -630	\\
W^{4,2}_2: 6	&	W^{7,1}_2: -21	&	W^{8,3}_4: -210	&	W^{9,3}_9: 28	&	W^{10,3}_5: 1512	\\
W^{4,2}_3: -8	&	W^{7,1}_3: 35	&	W^{8,3}_5: 336	&	W^{9,4}_4: 126	&	W^{10,3}_6: -2100	\\
W^{4,2}_4: 3	&	W^{7,1}_4: -35	&	W^{8,3}_6: -280	&	W^{9,4}_5: -504	&	W^{10,3}_7: 1800	\\
W^{4,3}_3: 4	&	W^{7,1}_5: 21	&	W^{8,3}_7: 120	&	W^{9,4}_6: 840	&	W^{10,3}_8: -945	\\
W^{4,3}_4: -3	&	W^{7,1}_6: -7	&	W^{8,3}_8: -21	&	W^{9,4}_7: -720	&	W^{10,3}_9: 280	\\
W^{4,4}_4: 1	&	W^{7,1}_7: 1	&	W^{8,4}_8: 35	&	W^{9,4}_8: 315	&	W^{10,3}_{10}: -36	\\
W^{5,1}_1: 5	&	W^{7,2}_2: 21	&	W^{8,4}_4: 70	&	W^{9,4}_9: -56	&	W^{10,4}_4: 210	\\
W^{5,1}_2: -10	&	W^{7,2}_3: -70	&	W^{8,4}_5: -224	&	W^{9,5}_8: -315	&	W^{10,4}_5: -1008	\\
W^{5,1}_3: 10	&	W^{7,2}_4: 105	&	W^{8,4}_6: 280	&	W^{9,5}_9: 70	&	W^{10,4}_6: 2100	\\
W^{5,1}_4: -5	&	W^{7,2}_5: -84	&	W^{8,4}_7: -160	&	W^{9,5}_5: 126	&	W^{10,4}_7: -2400	\\
W^{5,1}_5: 1	&	W^{7,2}_6: 35	&	W^{8,5}_8: -35	&	W^{9,5}_6: -420	&	W^{10,4}_8: 1575	\\
W^{5,2}_2: 10	&	W^{7,2}_7: -6	&	W^{8,5}_5: 56	&	W^{9,5}_7: 540	&	W^{10,4}_9: -560	\\
W^{5,2}_3: -20	&	W^{7,3}_3: 35	&	W^{8,5}_6: -140	&	W^{9,6}_8: 189	&	W^{10,4}_{10}: 84	\\
W^{5,2}_4: 15	&	W^{7,3}_4: -105	&	W^{8,5}_7: 120	&	W^{9,6}_9: -56	&	W^{10,5}_5: 252	\\
W^{5,2}_5: -4	&	W^{7,3}_5: 126	&	W^{8,6}_8: 21	&	W^{9,6}_6: 84	&	W^{10,5}_6: -1050	\\
W^{5,3}_3: 10	&	W^{7,3}_6: -70	&	W^{8,6}_6: 28	&	W^{9,6}_7: -216	&	W^{10,5}_7: 1800	\\
W^{5,3}_4: -15	&	W^{7,3}_7: 15	&	W^{8,6}_7: -48	&	W^{9,7}_8: -63	&	W^{10,5}_8: -1575	\\
W^{5,3}_5: 6	&	W^{7,4}_4: 35	&	W^{8,7}_8: -7	&	W^{9,7}_9: 28	&	W^{10,5}_9: 700	\\
W^{5,4}_4: 5	&	W^{7,4}_5: -84	&	W^{8,7}_7: 8	&	W^{9,7}_7: 36	&	W^{10,5}_{10}: -126	\\
W^{5,4}_5: -4	&	W^{7,4}_6: 70	&	W^{8,8}_8: 1	&	W^{9,8}_8: 9	&	W^{10,6}_8: 945	\\
W^{5,5}_5: 1	&	W^{7,4}_7: -20	&	W^{9,1}_1: 9	&	W^{9,8}_9: -8	&	W^{10,6}_9: -560	\\
W^{6,1}_1: 6	&	W^{7,5}_5: 21	&	W^{9,1}_2: -36	&	W^{9,9}_9: 1	&	W^{10,6}_{10}: 126	\\
W^{6,1}_2: -15	&	W^{7,5}_6: -35	&	W^{9,1}_3: 84	&	W^{10,1}_1: 10	&	W^{10,6}_6: 210	\\
W^{6,1}_3: 20	&	W^{7,5}_7: 15	&	W^{9,1}_4: -126	&	W^{10,1}_2: -45	&	W^{10,6}_7: -720	\\
W^{6,1}_4: -15	&	W^{7,6}_6: 7	&	W^{9,1}_5: 126	&	W^{10,1}_3: 120	&	W^{10,7}_8: -315	\\
W^{6,1}_5: 6	&	W^{7,6}_7: -6	&	W^{9,1}_6: -84	&	W^{10,1}_4: -210	&	W^{10,7}_9: 280	\\
W^{6,1}_6: -1	&	W^{7,7}_7: 1	&	W^{9,1}_7: 36	&	W^{10,1}_5: 252	&	W^{10,7}_{10}: -84	\\
W^{6,2}_2: 15	&	W^{8,1}_1: 8	&	W^{9,1}_8: -9	&	W^{10,1}_6: -210	&	W^{10,7}_7: 120	\\
W^{6,2}_3: -40	&	W^{8,1}_2: -28	&	W^{9,1}_9: 1	&	W^{10,1}_7: 120	&	W^{10,8}_8: 45	\\
W^{6,2}_4: 45	&	W^{8,1}_3: 56	&	W^{9,2}_2: 36	&	W^{10,1}_8: -45	&	W^{10,8}_9: -80	\\
W^{6,2}_5: -24	&	W^{8,1}_4: -70	&	W^{9,2}_3: -168	&	W^{10,1}_9: 10	&	W^{10,8}_{10}: 36	\\
W^{6,2}_6: 5	&	W^{8,1}_5: 56	&	W^{9,2}_4: 378	&	W^{10,1}_{10}: -1	&	W^{10,9}_9: 10	\\
W^{6,3}_3: 20	&	W^{8,1}_6: -28	&	W^{9,2}_5: -504	&	W^{10,2}_2: 45	&	W^{10,9}_{10}: -9	\\
\hline
\end{array}
$
}}

\bibliographystyle{IEEEtran}
\bibliography{references}
\end{document}